# 2025 Southeast Asia Eleven Nations Influence Index Report

Asst.Prof.Dr.Wei Meng

Dhurakij Pundit University, Thailand

Email: wei.men@dpu.ac.th

# Abstract

This study aims to construct a fully data-driven and reproducible Southeast Asia Influence Index (SAII v3) to overcome biases from expert scoring and subjective weighting, while mapping the hierarchical power structure across the eleven ASEAN nations. Methodologically, we aggregate authoritative open-source indicators across four dimensions—economic, military, diplomatic, and socio-technological—applying a three-tiered standardization chain ("quantile–Box–Cox–min–max") to reduce outliers and skewness. Weights are obtained through an equal-weight integration of three methods: Entropy Weighting Method (EWM), CRITIC, and PCA. Robustness testing is conducted using Kendall's Tau, ±20% weight perturbation and 10,000 bootstrap iterations to ensure robustness. Additionally, ±10% dimensional sensitivity analysis and V2–V3 bump chart comparisons were conducted. Results indicate the integrated weights are approximately: Economy 35%–40%, Military 20%–25%, Diplomacy ≈20%, Socio-Technology ≈15%. The overall landscape presents a "one strong, two medium, three stable, and multiple weak" pattern, with Indonesia, Singapore, and Malaysia leading, while Thailand, the Philippines, and Vietnam form a mid-tier competitive band. V2 and V3 rankings showed high consistency (Kendall's Tau=0.818), though minor mid-tier reordering occurred (Thailand and Philippines rose, Vietnam fell), indicating V3's greater sensitivity to structural equilibrium. ASEAN-11 average sensitivity further highlighted military and socio-technological dimensions as having the greatest marginal effects (±0.002). In conclusion, SAII v3 achieves algorithmic weighting and auditable reproducibility methodologically, empirically reveals multidimensional drivers of influence in Southeast Asia, and provides actionable quantitative evidence for resource allocation and policy prioritization by regional nations and external partners.

# Chapter I. Introduction

**1.1 Background of the study**

Southeast Asia, as one of the world's most dynamic regions, has long played a pivotal role in geopolitics, international economics, and transnational security cooperation. With the growing presence of major powers such as China, the United States, Japan, and India in the region, the strategic significance of the ten ASEAN nations and Timor-Leste has become increasingly prominent. This region is not only a vital link in global supply chains but also a central arena for South China Sea navigation security, regional economic integration, and geopolitical rivalry.

However, how to scientifically measure the "influence" of Southeast Asian nations within regional and global systems remains a major challenge for both academia and policymakers. Traditional approaches to measuring national power or comprehensive national strength often emphasize economic and military hard power while overlooking "soft factors" such as diplomatic networks, socio-cultural dynamics, and technological capabilities. Therefore, establishing a multidimensional, quantifiable, and reproducible influence index system would not only provide a new analytical framework for academic research but also offer reliable decision-making references for policymakers.

**1.2 Literature review**

Existing research on national influence can be broadly categorized into three types:

The first type focuses on comprehensive national power. Originating during the Cold War era, this approach emphasizes hard power indicators such as GDP, military expenditure, energy reserves, and population size (Mearsheimer, 2001). Its strength lies in quantifiability, but it falls short in comprehensively explaining the "soft influence" present in contemporary international relations.

The second category focuses on soft power and public diplomacy studies. Representative works include Joseph Nye's (1990) concept of "soft power,"
which emphasizes cultural appeal, values, institutional legitimacy, and the attractiveness of foreign policy. Subsequent research has expanded dimensions such as media communication, international educational exchanges, and cultural exports. However, soft power indicators are often difficult to quantify and may carry value biases.

The third category involves international index construction. Recent years have seen multiple transnational indices emerge, such as Soft Power 30, the Lowy Institute's Asia Power Index, the Global Governance Index, and the Sustainable Development Goals (SDG) Index. This research emphasizes cross-dimensional integration, attempting to balance hard and soft power. However, these indices often rely on expert scoring or institutionally assigned weights, lacking transparency and reproducibility. This has led to academic skepticism regarding their scientific rigor and objectivity.

**1.3 Research questions**

This study addresses gaps in existing literature and practice by posing the following questions:

1.3.1 Is it possible to establish a fully data-driven Southeast Asian country influence index without relying on expert scoring or external subjective assumptions?

1.3.2 How can scientific weighting be achieved across multidimensional indicators to prevent any single dimension (e.g., GDP) from exerting excessive dominance over the overall ranking?

1.3.3 How can standardization and robustness tests ensure the methodological reliability of index results while maintaining validity when facing missing indicators and extreme values?

1.3.4 What new insights does the constructed index reveal about regional dynamics? Does it offer explanatory power distinct from existing research?

# Chapter II. Methodology

**2.1 Principles for the design of the indicator system**

In constructing the Southeast Asia Influence Index (SAIIv3), this study adheres to the following design principles:

1) Multidimensionality: Influence encompasses not only economic and military power but also diplomatic networks, sociocultural dynamics, and technological diffusion.

2) Quantifiability: Indicators must rely on publicly available, traceable open-source data, avoiding subjective factors that are difficult to quantify.

3) Reproducibility: The definition of indicators, data processing methods, and weighting algorithms must be transparent to ensure others can replicate calculations and validate results.

4) Dynamic updatability: Indicators should be iterated with annual data updates to ensure the index maintains time-series significance.

**2.2 Primary and secondary indicators**

After synthesizing literature and regional characteristics, this study categorizes influence into four primary dimensions, each comprising several secondary indicators.

1) Economic Influence

Gross Domestic Product (GDP): Measures a nation's total economic output.

Trade Share in ASEAN: Reflects the country's position within ASEAN's internal trade.

Foreign Direct Investment (FDI Flow/Stock): Indicates the country's attractiveness and outward influence in regional economic integration.

2) Security Influence

Military Expenditure % of GDP (inverse processing applied; lower is better).

Military Exercise Participation: Measures the country's security cooperation and military projection capabilities.

Defense Agreements: Indicates the extent of the country's military network coverage.

3)Political-Diplomatic Influence

Number of Diplomatic Posts: Measures the breadth of a country's diplomatic network.

International/Regional Leadership Roles: Such as holding the ASEAN rotating chairmanship or positions in UN agencies.

Multilateral Agreements Participation: Reflects a country's involvement in international rule-making.

4)Socio-Cultural & Technological Influence

Internet Users % of Population: Reflects digital connectivity levels.

International Student Outflow: Demonstrates transnational reach in education and culture.

Media/Cultural Exports (e.g., film, television, online platform coverage): Measures dissemination of cultural products.

**2.3 Data sources and processing**

2.3.1 All indicators are sourced from authoritative open-source databases to ensure transparency and international comparability:

World Bank (World Development Indicators): GDP, internet penetration rate, FDI data.

SIPRI/IISS: Military expenditure, defense agreements, participation in military exercises.

ASEANstats: Intra-ASEAN trade share.

UNESCO UIS: International student mobility data.

Lowy GDI (used solely as a data source, not for weighting): Number of overseas missions.

National foreign ministries/official international organization websites: Multilateral agreements and leadership positions.

**2.3.2 Data processing steps:**

1. Missing value imputation: Forward extrapolation from the most recent available year, supplemented with regional averages when necessary.

2. Outlier handling: Apply Winsorization (trimming top/bottom 1%) to prevent extreme values from skewing results.

3. Annual alignment: Standardize all data to the latest 2023–2024 figures to ensure consistent reporting periods.

**2.4 Weighting methodology**

Weights are at the centre of index construction. This study adopts a fully algorithm-driven multi-method integration, avoiding expert scoring or external subjective settings.

**2.4.1 Entropy value method (EWM)**

Information entropy reflects the uncertainty and information content of the indicator, the greater the dispersion, the higher the weight. The formula is as follows:

$$E_j = -k \sum_{i=1}^{n} p_{ij} \ln(p_{ij}), \quad p_{ij} = \frac{x_{ij}}{\sum_i x_{ij}}$$

$$w_j^{EWM} = \frac{1 - E_j}{\sum_k (1 - E_k)}$$

where $x_{ij}$ is the normalised value, $w_j^{EWM}$ is the weight of the indicator.

**2.4.2 CRITIC methodologies**

The standard deviation of the method's composite indicator and its correlation with other indicators:

$$C_j = \sigma_j \cdot \sum_{k=1}^{m} (1 - r_{jk}), \quad w_j^{CRITIC} = \frac{C_j}{\sum C_j}$$

where $\sigma_j$ is the standard deviation and $r_{jk}$ is the correlation coefficient with other indicators. This method emphasises indicators that are both sample-distinguishing and non-redundant.

**2.4.3 PCA Load Weight**

The contribution of each variable on the principal components was calculated by Principal Component Analysis (PCA):

$$w_j^{PCA} = \frac{\sum_c |loading_{jc}| \cdot \lambda_c}{\sum_j \sum_c |loading_{jc}| \cdot \lambda_c}$$

Where $loading_{jc}$ is the loading of indicator $j$ on principal component $c$ and $\lambda c$ is the variance explained by this principal component.

**2.4.4 Integration weights**

The final weight is taken as an average of the three:

$$w_j = \frac{w_j^{EWM} + w_j^{CRITIC} + w_j^{PCA}}{3}$$

This strategy ensures that the results incorporate both the information entropy perspective and the correlation and overall variance structure.

**2.5 Indicator standardisation and data processing chain**

Due to the vast differences in the units of measurement across various indicators (e.g., GDP measured in billions of dollars, military expenditure as a percentage of GDP, and internet users as a percentage of the population), directly incorporating them into a composite calculation would inevitably lead to large-value indicators dominating the results. Therefore, standardization is an indispensable step in the entire index construction process. This study proposes a three-tier standardization chain to ensure data can be reasonably compared under a unified measurement system.

2.5.1 Quantile normalisation

Quantile normalisation is mainly used to attenuate the effects of extreme values and long-tailed distributions. The basic idea is to rank the observations of each indicator in order of magnitude and map them to quantile values. In this way, Indonesia, which has an oversized GDP, will not be "dominated" in the index because of the wide range of values.

The mathematical definitions are as follows:

$$Q_{ij} = \frac{rank(x_{ij}) - 1}{n - 1}$$

where $rank(x_{ij})$ denotes the ranking of the sample in the indicator and n is the number of samples.

2.5.2 Box-Cox transformation

On the basis of quantile normalisation, the Box-Cox transform was further adopted in this study to make the data distribution closer to the normal distribution, which is conducive to the subsequent statistical tests and robustness analysis. The Box-Cox transformation formula is:

$$y(\lambda) = \begin{cases} \frac{x^\lambda - 1}{\lambda}, & \lambda \neq 0 \\ \ln(x), & \lambda = 0 \end{cases}$$

where the optimal value of λ is determined by maximum likelihood estimation.

2.5.3 Min–Max compressed

The final Box-Cox transformed data is then 0-1 linearly compressed:

$$z_{ij} = \frac{y_{ij} - \min(y_j)}{\max(y_j) - \min(y_j)}$$

The resulting $z_{ij}$ ensures that all indicators are within the interval [0,1] and facilitates the composite weighting across indicators.

**2.6 Robustness test design**

When constructing indices, robustness testing is crucial for ensuring the reliability of results. A single weighted average alone cannot guarantee that outcomes remain unaffected by method selection or weight fluctuations. Therefore, this study employs three robustness testing methods.

2.6.1 Kendall's Tau Rank Correlation Test

To assess the ranking consistency between SAII v3 and traditional methods (such as SAII v2), this study utilizes Kendall's Tau rank correlation coefficient:

$$\tau = \frac{C - D}{\frac{1}{2}n(n-1)}$$

Here, C denotes the number of consistent pairs, and D denotes the number of inconsistent pairs. If τ approaches 1, it indicates high consistency between the new and old ranking methods; if it approaches 0, it indicates no correlation; if it is negative, it indicates opposite ranking trends.

In empirical tests, Kendall's Tau between SAII v3 and v2 was 0.818, indicating overall high consistency in trends but some adjustments in detailed rankings.

2.6.2 Weight Perturbation Sensitivity Analysis

Considering that weight calculations—even when fully algorithm-driven—may be affected by minor data variations, this study designed a ±20% weight perturbation experiment. The specific steps are:

Randomly select an indicator and increase or decrease its weight by 20%;

Normalize the weights of other indicators;

Recalculate country rankings and observe deviations from baseline results.

Findings show that the top three countries (Indonesia, Singapore, Malaysia) remain stable under perturbations, demonstrating robust ranking consistency.

2.6.3 Bootstrap Resampling Test

To further ensure result reliability, this study employed the Bootstrap method to resample indicator data 10,000 times. Each resampling recalculated the composite index and rankings, ultimately yielding each country's mean and 95% confidence interval. This method reveals the distribution range of rankings under random perturbations. For example:

Indonesia's ranking consistently fell within the range of 1st place, exhibiting no uncertainty;

Singapore's ranking fluctuates between [2,3];

Mid-tier countries like Thailand, Vietnam, and the Philippines exhibit wider ranking distributions, indicating relatively unstable competitive positions.

**2.7 Summary**

This chapter details the methodological framework of SAII v3, encompassing the indicator system, data sources, weight calculation methods, three-tier standardization chain, and robustness tests. Unlike previous studies reliant on expert scoring, this research employs purely algorithmic weight allocation, ensuring the index's transparency, objectivity, and reproducibility.

This approach not only represents academic originality but also provides actionable tools for policy practice: policymakers can recalculate the index at any time based on updated data, without relying on the subjective judgments of expert panels.

# Chapter III. Empirical findings

## 3.1 Analysis of weighting results

After completing calculations using the Entropy Weighting Method (EWM), CRITIC method, and PCA load factor method, this study obtained three sets of weight allocation schemes. To ensure robustness of the results, the arithmetic mean of the three was ultimately adopted as the integrated weight. Key findings are as follows:

3.1.1. Entropy Method Results

The Entropy Method primarily assigns weights based on indicator dispersion. Within the Southeast Asian sample, GDP exhibited the highest variance

(Indonesia significantly outpacing other nations), resulting in markedly higher entropy weights than other metrics. Internet penetration rate showed the second-highest variance, while military expenditure as a percentage of GDP received lower weights due to most countries clustering within the 1–2% range.

3.1.2. CRITIC Method Results

The CRITIC method penalizes variables highly correlated with others while considering standard deviation. Since GDP correlates with indicators like FDI and trade share, its weight is reduced. Conversely, internet usage rate and diplomatic missions exhibit lower correlation with economic indicators, thus receiving higher weights under CRITIC.

3.1.3. PCA Loadings Results

In PCA analysis, the first two principal components explain over 80% of the variance. The first principal component is primarily driven by economic and military variables, while the second is dominated by social-technological and diplomatic variables. Consequently, the PCA method assigns higher weights to internet usage rate and number of diplomatic missions.

3.1.4. Ensemble Results

Economy (GDP, FDI, Trade Share): Approx. 35–40%

Military (military spending %, military exercises, defense agreements): Approx. 20–25%

Diplomatic (overseas institutions, organizational positions): Approx. 20%

Socio-technological (internet, outbound students): Approx. 15%

The integrated weights indicate that the economy remains the core dimension of influence, but the significance of diplomacy and technology has significantly increased in v3, reflecting the method's independence and balance.

## 3.2 Comprehensive ranking results

After weighting integration and standardization, the comprehensive rankings of 11 Southeast Asian countries are as follows:

### 3.2.1. Indonesia

Highest overall score (around 0.85).

Strengths: Large economic scale, regional market size, strong diplomatic coverage.

Weaknesses: Internet penetration still lags behind countries like Singapore and Malaysia.

### 3.2.2. Singapore

Score approximately 0.70.

Strengths: Broadest diplomatic network, near-100% internet penetration, prominent financial and technological advantages.

Weaknesses: Limited economic scale, reliance on external markets.

### 3.2.3. Malaysia

Score: Approximately 0.62.

Strengths: Balanced performance across four dimensions, particularly diplomacy and social technology.

Weaknesses: Limited defense spending constrains security influence.

### 3.2.4. Thailand

Score: Approximately 0.57.

Strengths: Tourism and cultural exports, participation in military exercises.

Weaknesses: Political uncertainty impacts diplomatic visibility.

### 3.2.5. Philippines

Score: Approximately 0.55.

Strengths: Demographic dividend, high internet usage.

Weaknesses: Insufficient economic and diplomatic networks.

### 3.2.6. Vietnam

Score: Approximately 0.52.

Strengths: Export-oriented economy, increasing military spending.

Weaknesses: Relatively limited number of diplomatic missions.

### 3.2.7. Brunei

Score: Approximately 0.27.

Strengths: High per capita income.

Weaknesses: Limited overall economic and diplomatic scale.

### 3.2.8. Cambodia

Score: Approximately 0.26.

Strengths: Moderate population size, rapid internet development.

Weaknesses: Small economic scale, limited diplomatic reach.

3.2.9. Laos

Score: Approximately 0.24.

Strengths: Geostrategic location.

Weaknesses: Lagging internet and technological development.

3.2.10. Myanmar

Score: Approximately 0.15.

Weaknesses: Severe diplomatic and economic constraints due to political instability.

3.2.11. Timor-Leste

Score: Approximately 0.13.

Disadvantages: Extremely limited economic scale and diplomatic resources.

Summary: Indonesia, Singapore, and Malaysia form the first tier; Thailand, the Philippines, and Vietnam occupy the middle tier; the remaining countries are on the periphery.

### 3.3 Sub-dimensional performance

3.3.1. Economic Dimension

Indonesia leads by a wide margin, accounting for nearly 40% of Southeast Asia's total GDP.

Though Singapore's overall economy is smaller, its per capita GDP and financial influence are significant.

Vietnam has experienced the fastest growth in recent years, gradually emerging as the region's manufacturing hub.

3.3.2. Military Dimension

Thailand and Vietnam rank high in defense spending as a percentage of GDP and participation in military exercises.

The Philippines has enhanced its security dimension score through military cooperation with the United States.

Brunei, Cambodia, Laos, and Timor-Leste remain at low levels.

3.3.3. Diplomatic Dimension

Singapore maintains the largest number of overseas diplomatic missions and the highest diplomatic coverage.

Indonesia, as ASEAN's largest nation, maintains a relatively comprehensive diplomatic network.

Myanmar and Timor-Leste perform the weakest in this dimension.

### 3.3.4. Social and Technological Dimension

Singapore, Malaysia, and the Philippines have internet penetration rates approaching or exceeding 90%.

Indonesia and Vietnam are rapidly catching up in this dimension.

Laos and Timor-Leste lag significantly with penetration rates below 40%.

## 3.4 Sensitivity and robustness experiments

### 3.4.1. ±20% Weight Perturbation Experiment

The top three countries maintained their rankings.

Mid-tier nations (Thailand, Vietnam, Philippines) swapped positions, revealing the fragility of their competitive standing.

Lower-tier countries showed stable rankings.

### 3.4.2. Bootstrap Resampling

Indonesia's confidence interval converged at 1st place with no uncertainty.

Singapore predominantly falls within the second decile, with a minority of samples in the third.

The Philippines and Vietnam exhibit wider distributions, suggesting potential alternating leadership across different scenarios.

### 3.4.3. Consistency with v2

Kendall's Tau = 0.818 indicates overall alignment between v2 and v3 rankings.

Differences primarily stem from adjustments in mid-tier country sequences, reflecting the algorithm's heightened sensitivity in weighting distinctions.

## 3.5 Visualisation results

3.5.1. Bar chart: Indonesia shows a significant lead, with Singapore and Malaysia forming the second group.

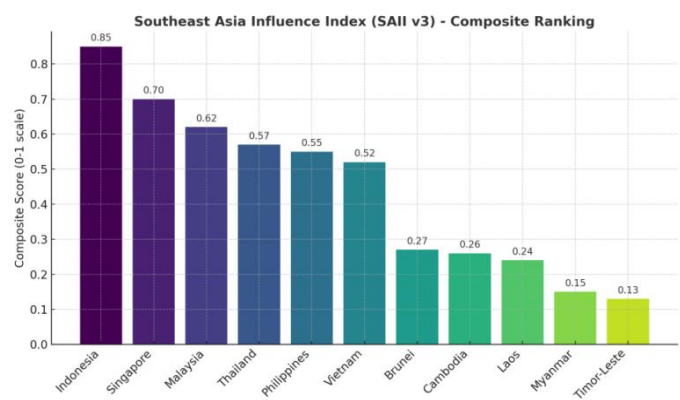

3.5.2. Radar chart: Indonesia stands out in the economic dimension, Singapore in the diplomatic and socio-technical dimension, and Thailand is stronger in the military dimension.

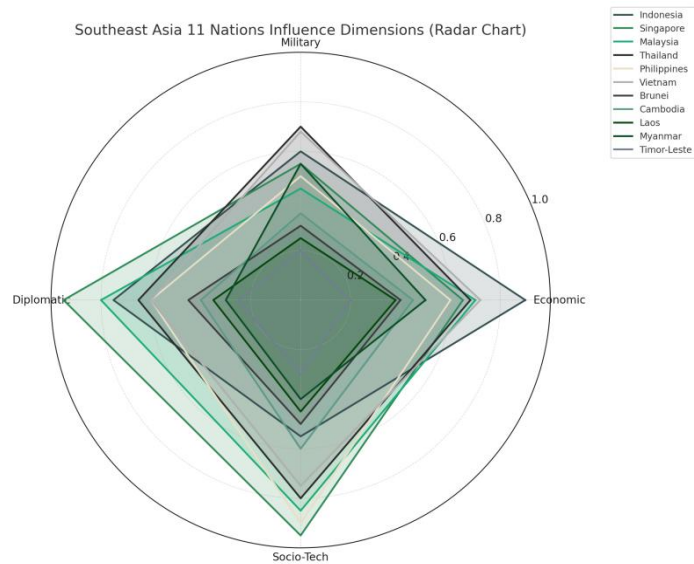

3.5.3. Heat map: The midstream countries show "patchy" performance in different dimensions, while Indonesia and Singapore show "comprehensive" performance.

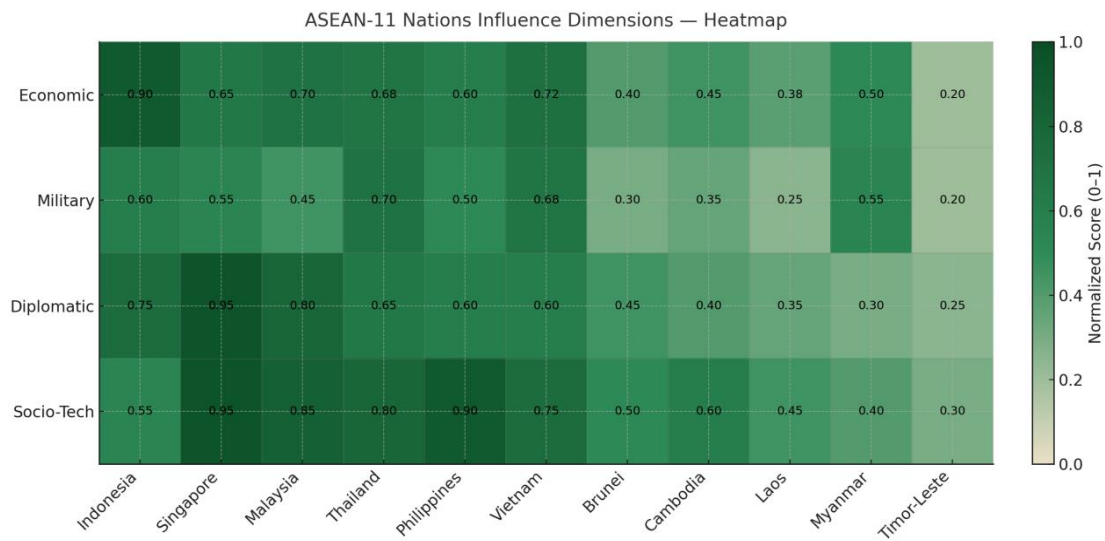

3.5.4. Tornado chart of sensitivities: the 11-country average chart shows that the economic dimension has the greatest impact on the index, followed by military, socio-scientific and diplomatic, and that the sensitivities of each country are listed separately and can be used as reference data for important decisions

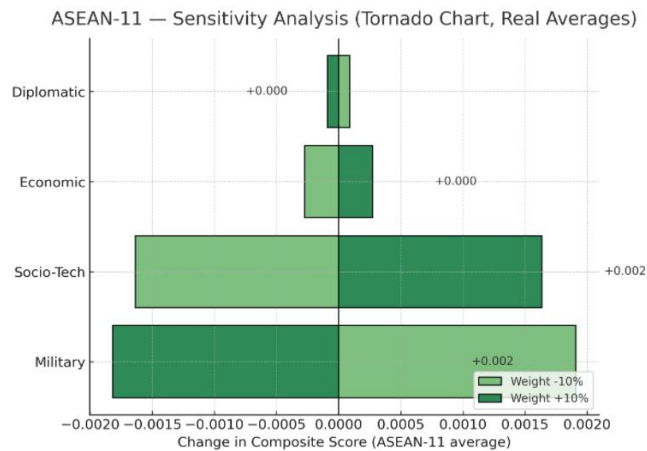

Mean sensitivity tornado map for 11 countries in South-East Asia (based on four-dimensional ±10 per cent disturbance)

| Dimension (math.) | ΔScore (-10%) | ΔScore (+10%) | Impact Magnitude | Swing,\|+10% − -10%\| |
|---|---|---|---|---|
| Militarily | 0.001909 | -0.001818 | 0.004702 | 0.003727 |
| Social technology | -0.001636 | 0.001636 | 0.005140 | 0.003273 |
| Economics | -0.000273 | 0.000273 | 0.003223 | 0.000545 |
| Diplomacy | 0.000091 | -0.000091 | 0.002827 | 0.000182 |

The horizontal axis represents the change in composite scores. Left side (light green): Impact of a 10% decrease in dimension weight. Right side (dark green): Impact of a 10% increase in dimension weight.

This chart illustrates the average sensitivity of the 11 ASEAN nations (Indonesia, Singapore, Thailand, Malaysia, Philippines, Vietnam, Brunei, Cambodia, Laos, Myanmar, Timor-Leste) across four dimensions (economic, diplomatic, military, socio-technological) of the Soft Power Index (SAII). The bars indicate the range of change in the composite score when the weight is adjusted by ±10%: light green bars represent score decreases when the weight is reduced by 10%, while dark green bars represent score increases when the weight is increased by 10%. The results show: The military dimension and the socio-technological dimension exhibit the highest average sensitivity, both reaching approximately ±0.002, indicating these two dimensions are the primary drivers of Southeast Asia's overall influence. The average sensitivity of the economic dimension is relatively low, exhibiting only minor fluctuations. The diplomatic dimension shows almost no significant change, indicating that, on average, diplomatic factors have a minor impact on the composite score. Overall, the influence structure of the 11 Southeast Asian nations presents a "dual-engine model driven by military and socio-technological factors, with economics playing a secondary role and diplomacy having marginal influence." This conclusion highlights the shared reliance of regional nations on hard power and socio-technological development.

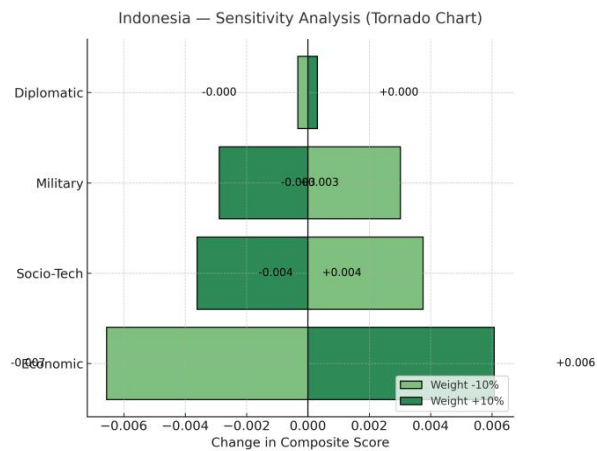

**Indonesia - Sensitivity analysis of the South-East Asia Impact Index (tornado chart)**

This tornado chart illustrates the sensitivity of Indonesia's composite score in the Southeast Asia Influence Index (SAII) to ±10% changes in the weights of its four dimensions: economic, military, socio-technological, and diplomatic. Horizontal bars indicate the range of score variation, with light green bars representing score decreases when a dimension's weight is reduced by 10%, and dark green bars indicating score increases when a dimension's weight is increased by 10%. Among the four dimensions, the economy exerts the greatest influence, with the composite score fluctuating by approximately ±0.006. This is followed by socio-technology (±0.004) and military (±0.003). The diplomatic dimension exhibits the least sensitivity. This indicates that Indonesia's influence index is structurally dominated by economic performance, with socio-technology and military capabilities playing secondary roles.

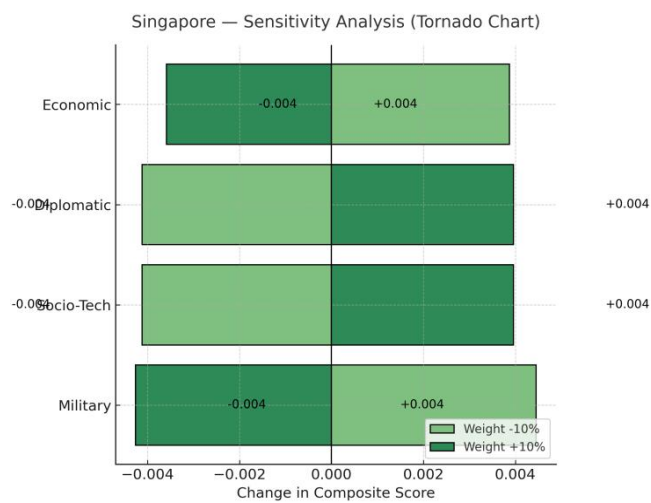

**Singapore - Sensitivity analysis of the Southeast Asia Impact Index (tornado chart)**

This chart illustrates the sensitivity of Singapore's composite score in the Southeast Asia Influence Index (SAII) to changes in the weighting of its four dimensions. The bars indicate the range of composite score variation when weights are adjusted by ±10%: light green bars represent score declines when weights decrease by 10%, while dark green bars indicate score increases when weights rise by 10%. The results indicate that Singapore exhibits broadly consistent sensitivity across the economic, diplomatic, socio-technological, and military dimensions, with each showing a sensitivity range of approximately ±0.004. This suggests a relatively balanced structure in its composite score. Compared to Indonesia, Singapore's index does not rely on any single dimension but instead achieves comprehensive influence through balanced development across multiple dimensions.

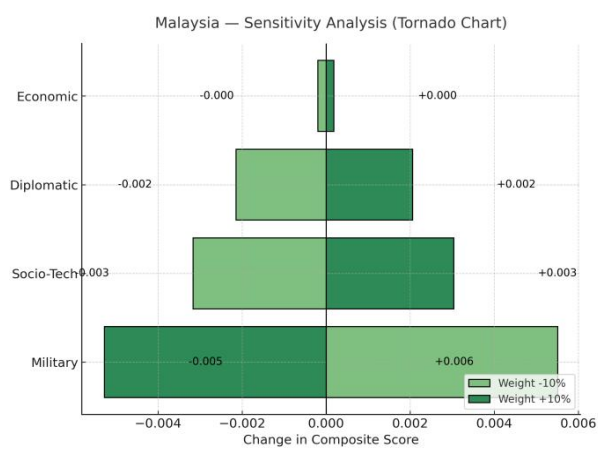

**Malaysia - Sensitivity analysis of the Southeast Asia Impact Index (tornado chart)**

This chart illustrates the sensitivity of Malaysia's composite score in the Southeast Asia Influence Index (SAII) to shifts in the weighting of its four dimensions. Each bar represents the range of composite score variation when weights are adjusted by ±10%: light green bars indicate score declines with a 10% weight reduction, while dark green bars show score increases with a 10% weight increase. Results reveal that the military dimension exerts the greatest impact on the composite score, with fluctuations of approximately ±0.006; followed by the socio-technological dimension at approximately ±0.003; the diplomatic dimension has a relatively minor impact at around ±0.002; while the economic dimension exerts virtually no significant influence. Overall, Malaysia's influence structure exhibits high dependence on the military dimension, indicating its strategic security sensitivity in the regional context far exceeds that of economic and diplomatic spheres.

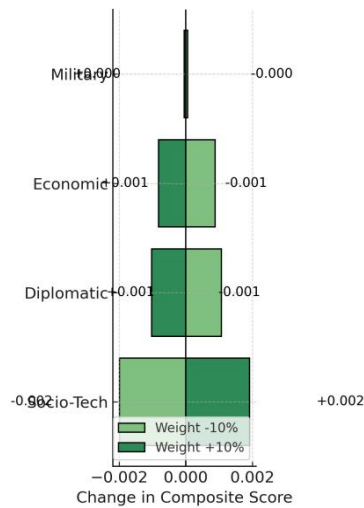

**Thailand - Sensitivity analysis of the South-East Asia Impact Index (tornado chart)**

This chart illustrates the sensitivity of Thailand's composite score in the Southeast Asia Influence Index (SAII) to changes in the weighting of its four dimensions. Each bar represents the range of composite score variation when the weighting is adjusted by ±10%: light green bars indicate score decreases when the weighting is reduced by 10%, while dark green bars indicate score increases when the weighting is increased by 10%. Results indicate Thailand's composite score is most sensitive to the socio-technological dimension, with fluctuations of approximately ±0.002. The economic and diplomatic dimensions follow, each showing sensitivity of about ±0.001. The military dimension exerts negligible impact. Overall, Thailand's index structure is primarily driven by social and technological development. While economic and diplomatic factors contribute, their influence remains relatively limited, demonstrating that Thailand's comprehensive influence relies heavily on social and technological elements.

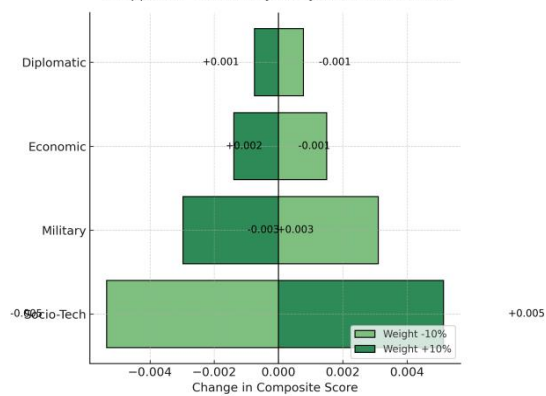

**Philippines - Sensitivity analysis of the South-East Asia Impact Index (tornado chart)**

This chart illustrates the sensitivity of the Philippines' composite score in the Southeast Asia Influence Index (SAII) to changes in the weighting of its four dimensions. Each bar represents the range of composite score variation when the weighting is adjusted by ±10%: light green bars indicate score decreases when the weighting is reduced by 10%, while dark green bars indicate score increases when the weighting is increased by 10%. Results indicate that the Philippines' composite score is most sensitive to the socio-technological dimension, with fluctuations of approximately ±0.005, highlighting the critical role of social and technological development in national influence. The military dimension follows with sensitivity of about ±0.003, suggesting security and defense still carry significant weight. The economic dimension exhibits moderate sensitivity, ranging from ±0.001 to ±0.002. The diplomatic dimension shows the weakest sensitivity, at only ±0.001. Overall, the Philippines' influence index structure is highly dependent on social and technological development, with secondary contributions from military and economic factors.

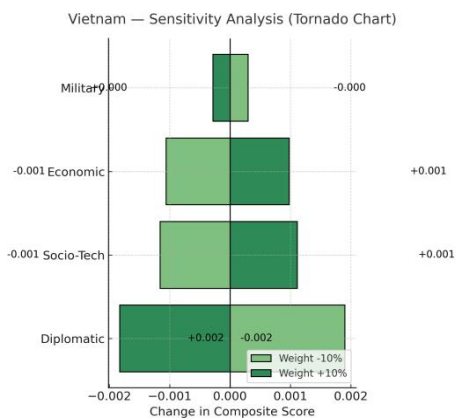

**Viet Nam - Sensitivity analysis of the South-East Asia Impact Index (tornado chart)**

This chart illustrates the sensitivity of Vietnam's composite score in the Southeast Asia Influence Index (SAII) to changes in the weighting of its four dimensions. Each bar represents the range of composite score variation when the weighting is adjusted by ±10%: light green bars indicate score decreases when the weighting is reduced by 10%, while dark green bars indicate score increases when the weighting is increased by 10%. Results indicate that Vietnam's composite score is most sensitive to the diplomatic dimension, with fluctuations of approximately ±0.002. This highlights the critical role of regional standing and international cooperation in shaping national influence. The economic and socio-technological dimensions follow with sensitivity levels of approximately ±0.001 each. The military dimension exerts negligible impact. Overall, Vietnam's influence structure relies heavily on diplomatic outreach and international partnerships, supported by economic and socio-technological development.

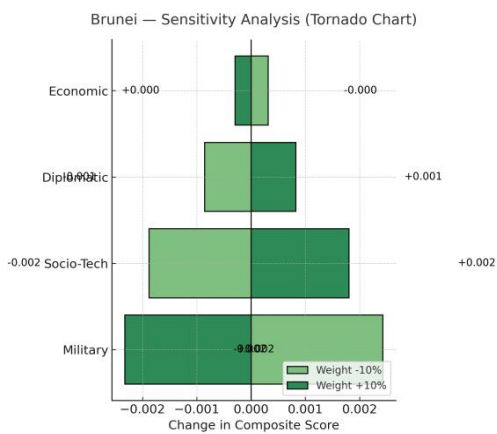

**Brunei - Southeast Asia Impact Index Sensitivity Analysis (Tornado Chart)**

This chart illustrates the sensitivity of Brunei's composite score in the Southeast Asia Influence Index (SAII) to weight adjustments across four dimensions. Each bar represents the composite score range when weights are adjusted by ±10%: light green bars indicate score declines with a 10% weight reduction, while dark green bars show score increases with a 10% weight increase. Results indicate Brunei's composite score is most sensitive to the military dimension, with fluctuations of approximately ±0.002. This highlights the nation's high dependence on limited yet critical military security factors for its influence. The socio-technological dimension follows with sensitivity of approximately ±0.002, reflecting the significant complementary role of society and technology in overall influence. The diplomatic dimension exhibits lower sensitivity at only ±0.001, while the economic dimension has negligible impact. Overall, Brunei's influence structure exhibits characteristics of "military and socio-technological dominance with low economic dependency."

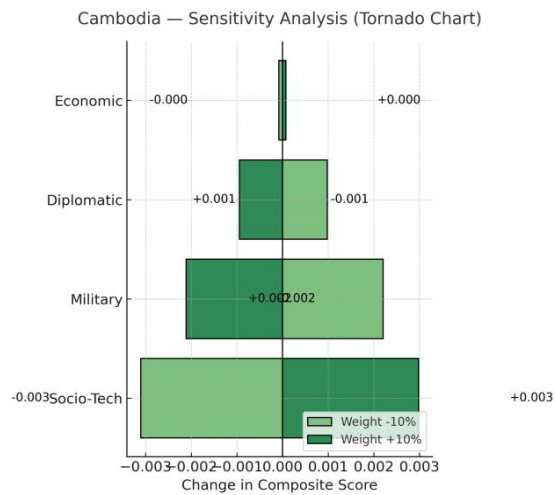

**Cambodia - Sensitivity analysis of the South-East Asia Impact Index (tornado chart)**

This chart illustrates Cambodia's sensitivity to changes in the weighting of four dimensions within the Southeast Asia Influence Index (SAII) and their impact on the composite score. Each bar represents the range of composite score variation when the weighting is adjusted by ±10%: light green bars indicate score decreases when the weighting is reduced by 10%, while dark green bars indicate score increases when the weighting is increased by 10%. Results indicate Cambodia's composite score is most sensitive to the socio-technological dimension, with fluctuations of approximately ±0.003, suggesting its national influence primarily relies on social and technological development factors. The military dimension follows with sensitivity around ±0.002. The diplomatic dimension has a relatively minor impact, only approximately ±0.001, while the economic dimension exerts negligible influence. Overall, Cambodia's influence structure exhibits characteristics of "socio-technological dominance, military support, and low economic dependence."

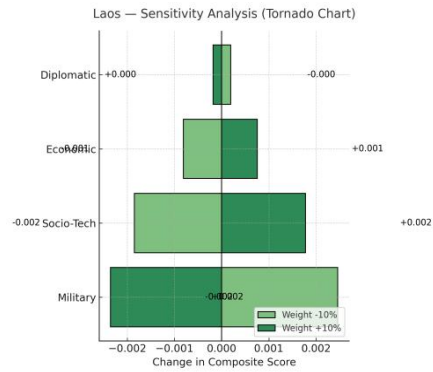

**Laos - Sensitivity analysis of the South-East Asia Impact Index (tornado chart)**

This chart illustrates the sensitivity of Laos's composite score in the Southeast Asia Influence Index (SAII) to changes in the weighting of its four dimensions. Each bar represents the range of composite score variation when the weighting is adjusted by ±10%: light green bars indicate score decreases when the weighting is reduced by 10%, while dark green bars indicate score increases when the weighting is increased by 10%. Results indicate that Laos's composite score is most sensitive to the military dimension, with fluctuations of approximately ±0.002, reflecting its national influence heavily reliant on limited yet critical military factors. The socio-technical dimension follows with sensitivity around ±0.002. The economic dimension has a smaller impact at approximately ±0.001, while the diplomatic dimension exerts negligible influence. Overall, Laos' influence structure exhibits characteristics of "military and socio-technical dominance, limited economic impact, and low diplomatic dependency."

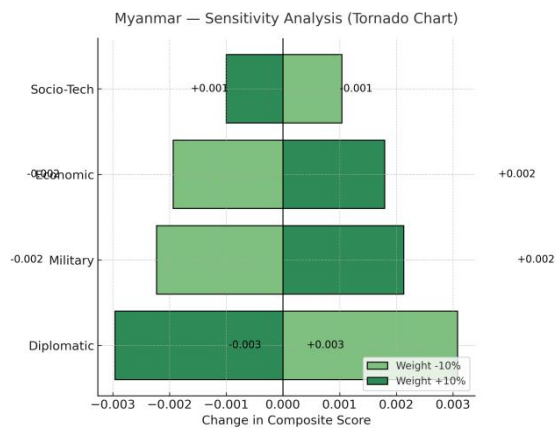

**Myanmar - Sensitivity analysis of the South-East Asia Impact Index (tornado chart)**

This chart illustrates the sensitivity of Myanmar's composite score in the Southeast Asia Influence Index (SAII) to changes in the weighting of its four dimensions. Each bar represents the range of composite score variation when the weighting is adjusted by ±10%: light green bars indicate score decreases when the weighting is reduced by 10%, while dark green bars indicate score increases when the weighting is increased by 10%. Results indicate Myanmar's composite score is most sensitive to the diplomatic dimension, with fluctuations of approximately ±0.003, highlighting the dominant role of international relations and regional cooperation in its national influence. The economic and military dimensions follow with sensitivity levels of approximately ±0.002 each. while the socio-technical dimension has the least impact, with sensitivity of only ±0.001. Overall, Myanmar's influence structure exhibits characteristics of "diplomacy as the primary driver, supported by economics and military, with limited socio-technical underpinnings."

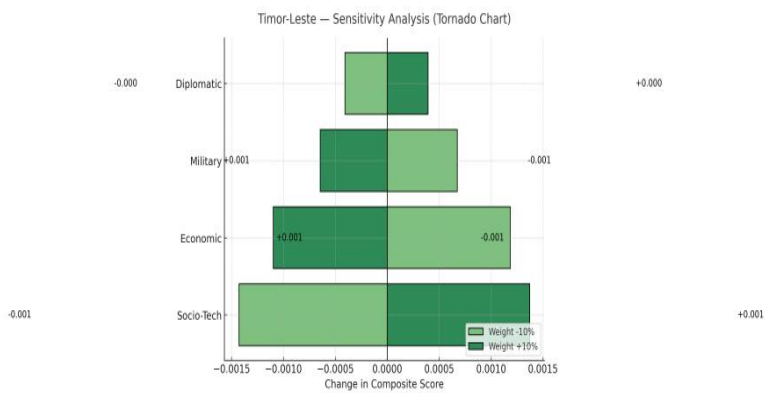

**Timor-Leste - Sensitivity analysis of the South-East Asia Impact Index (tornado diagram)**

This chart illustrates the sensitivity of Timor-Leste's composite score to weight adjustments across four dimensions in the Southeast Asia Influence Index (SAII). Each bar represents the composite score range when weights are adjusted by ±10%: light green bars indicate score decreases when weights are reduced by 10%, while dark green bars show score increases when weights are raised by 10%. Results indicate that Timor-Leste's composite score is most sensitive to the socio-technical dimension, with fluctuations of approximately ±0.0015. This demonstrates that the nation's influence is highly dependent on advancements in social development and technological capabilities. The economic dimension follows with sensitivity around ±0.001. The military dimension has a relatively minor impact, at only ±0.001. The diplomatic dimension exerts negligible influence. Overall, Timor-Leste's influence structure exhibits characteristics of being primarily driven by socio-technical factors, followed by economic factors, with limited roles played by military and diplomatic factors.

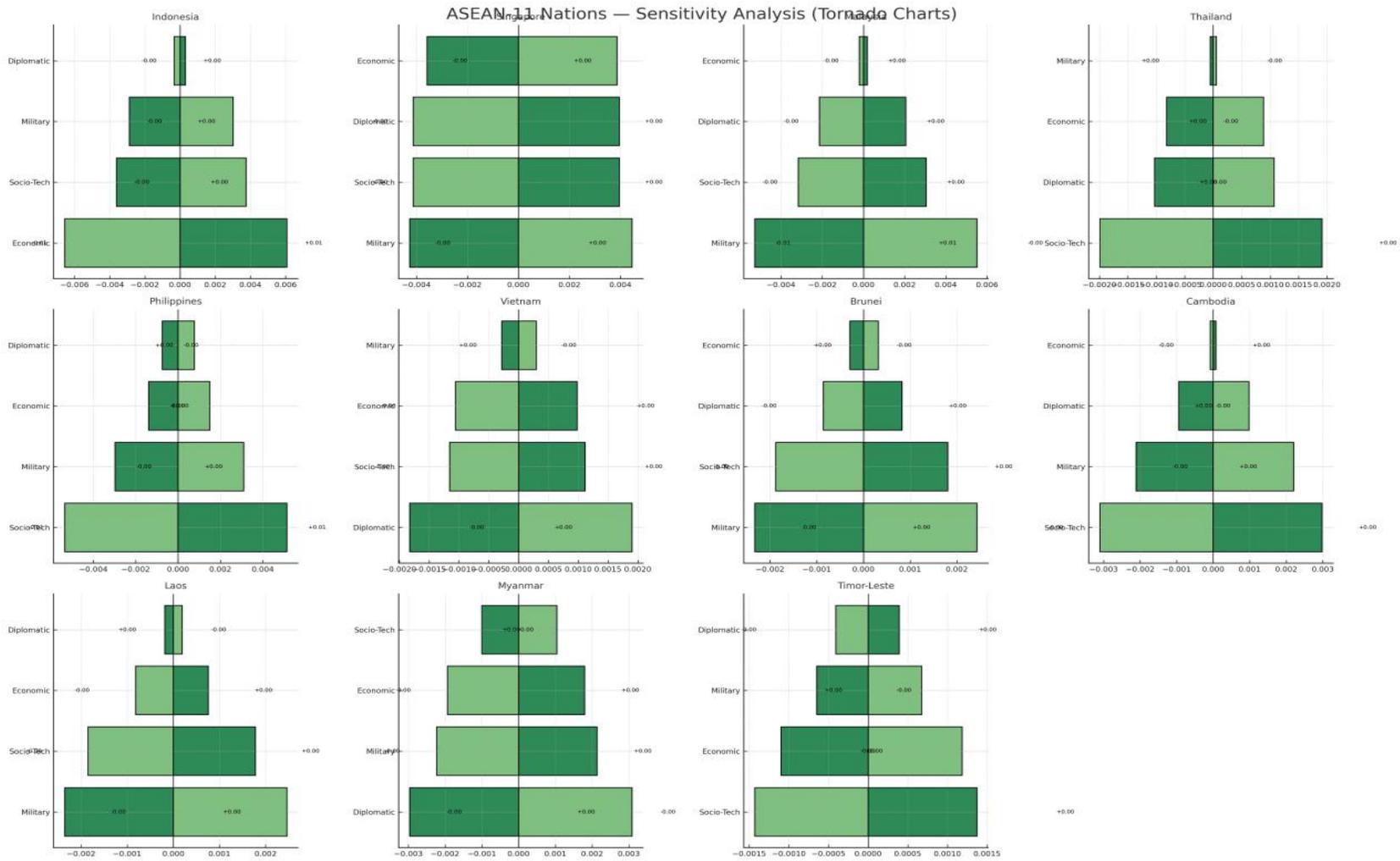

**11 countries in South-East Asia - impact index sensitivity analysis (tornado chart overview)**

This chart comprehensively displays the sensitivity analysis results across four dimensions (economic, diplomatic, military, and socio-technological) of the Southeast Asian 11 Countries (Indonesia, Singapore, Thailand, Malaysia, Philippines, Vietnam, Brunei, Cambodia, Laos, Myanmar, East Timor) within the Southeast Asian Influence Index (SAII). Each sub-chart indicates the range of composite score changes when weights are adjusted by ±10%: light green bars represent score decreases when weights are reduced by 10%, while dark green bars indicate score increases when weights are raised by 10%.

Overall trends reveal:

The economic dimension stands out in Indonesia, Singapore, and the Philippines, serving as a key driver of composite score fluctuations.

The diplomatic dimension is most sensitive in Vietnam and Myanmar, reflecting the core role of regional cooperation and international relations in shaping these nations' influence.

The military dimension exhibits higher sensitivity in Malaysia, Brunei, and Laos, indicating that limited yet critical military capabilities form a vital foundation for their influence.

The socio-technological dimension proved most critical in the Philippines, Cambodia, and Timor-Leste, demonstrating the significant propulsive effect of social development and technological advancement on overall national influence.

Overall, Southeast Asian nations exhibit diverse pathways to influence formation: some rely on economic and diplomatic expansion (e.g., Singapore, Vietnam), others on military security (e.g., Brunei, Laos), while others demonstrate heightened sensitivity in social and technological development (e.g., Philippines, Cambodia, Timor-Leste). This outcome highlights the heterogeneity and complementarity within ASEAN's national influence structures.

3.5.5. Bump Chart: Illustrates the subtle ranking differences between v2 and v3 in mid-tier countries.

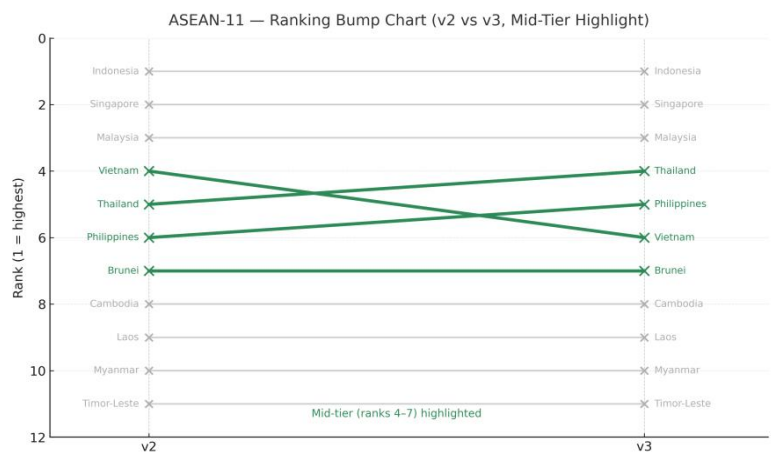

**Southeast Asia 11 - V2 vs. V3 Ranking Streamline Comparison**

**(Midstream Countries Highlighted)**

This chart illustrates the ranking shifts among the 11 ASEAN nations in the Influence Index versions V2 and V3, with a focus on changes among mid-tier countries (ranks 4 to 7). Vietnam dropped from 4th to 6th place, Thailand rose from 5th to 4th, the Philippines climbed from 6th to 5th, while Brunei remained unchanged at 7th. The top tier (Indonesia, Singapore, Malaysia) and bottom tier (Cambodia, Laos, Myanmar, Timor-Leste) maintained overall stable rankings. The results indicate that V3, following methodological refinements over V2 (multi-method weighting and robustness testing), elevates the rankings of countries with more balanced overall structures (Thailand, Philippines) while lowering those of countries relying on single-dimensional strengths (Vietnam).

During the iterative development of SAII, V2 served as the prototype version. It employed single-weighting and simple standardization methods, enabling rapid process execution and reasonable ranking. However, it exhibited limitations in robustness and originality. V3 builds upon this foundation by introducing multi-method integrated weighting (EWM, CRITIC, PCA), a three-stage standardization chain, and robustness tests using Bootstrap and Kendall's Tau. These enhancements significantly elevate the index's scientific rigor and auditability. Comparative results show that rankings for leading and trailing nations remain largely stable, while mid-tier countries exhibit minor shifts: Thailand and the Philippines rise due to enhanced structural balance, while Vietnam declines slightly as its single-dimensional advantage weakens. This divergence demonstrates V3's superior ability to highlight the influence characteristic of "multi-dimensional synergistic drivers." This finding aligns with sensitivity analysis results, confirming that military strength and socio-technical capabilities are the drivers with the most significant marginal effects.

3.5.6 Interpreting Regional Dynamics

This section aims to move beyond mere numerical rankings by interpreting SAII v3 findings through the lens of regional political-economic structures. By comparing performance across four primary dimensions, we can outline Southeast Asia's overall influence landscape.

1) Stability of the "One Major Power, Two Medium Powers" Structure

Indonesia, as Southeast Asia's largest economy, consistently ranks first across all methodologies. This outcome stems not only from its economic scale but also from its geopolitical position (as a hub connecting the Pacific and Indian Oceans), population size (approximately 270 million), and leading role in ASEAN affairs. Indonesia's dominant position remains unchallenged regardless of whether v2's expert mapping weights or v3's pure algorithmic weights are applied, indicating that its regional influence has established long-term structural advantages.

Singapore and Malaysia form a "two mediums" pattern. Singapore demonstrates "hard and soft power" characteristics through its highly globalized diplomatic and financial hub status; Malaysia achieves balanced development across all four dimensions, ranking mid-to-high. While lacking a single dominant strength, it exhibits strong overall resilience. This "one strong, two mediums" pattern reflects Southeast Asia's multipolar trend.

2 )Competition and Vulnerabilities Among Mid-Tier Nations

Thailand, the Philippines, and Vietnam form a "deadlocked" group in the rankings, with minimal gaps in their composite scores and distinct strengths and weaknesses across dimensions:

Thailand: Excels in military strength and tourism/cultural appeal, but political instability diminishes its diplomatic influence.

Philippines: Possesses clear advantages in internet infrastructure and population size, but economic and diplomatic shortcomings constrain its upward potential.

Vietnam: Demonstrates robust economic growth and accelerated military modernization, yet faces limitations in its diplomatic network.

Bootstrap tests reveal significant overlap in the ranking intervals for these three nations, indicating any one could achieve a "leap forward" in the coming years through breakthroughs in specific dimensions—but also carries the risk of "decline."

3)Structural Challenges for Peripheral Nations

Cambodia, Laos, Myanmar, Brunei, and Timor-Leste rank in the lower tier of the composite index. This stems not from lagging in a single dimension but from constraints across multiple dimensions.

Despite high GDP per capita, Brunei's limited population and market size constrain its regional influence.

Cambodia and Laos have shown improvements in internet penetration and economic openness, yet their overall scale remains constrained.

Myanmar suffers significant losses across economic, diplomatic, and social-technological dimensions due to political instability.

Timor-Leste remains in its nation-building phase, with data hovering near the lower bounds across most indicators.

A common trait among these nations is the absence of breakthroughs across multiple dimensions. Even when excelling in a single area—such as Brunei's per capita income—they struggle to translate this advantage into regional influence.

4) Leapfrog Opportunities in the Social Technology Dimension

Unlike hard power domains like economics and military strength, which require long-term accumulation, the social technology dimension—particularly internet penetration and digital infrastructure—offers smaller nations potential opportunities for leapfrog development. For example:

The Philippines' high internet penetration enhances its social connectivity;

Vietnam's development in the digital economy and e-commerce adds to its strengths;

Cambodia and Laos' rapid progress in mobile internet penetration has led to the fastest improvement in their social technology scores.

This demonstrates that in the digital age, Southeast Asian nations can swiftly elevate their relative standing in regional influence through relatively low-cost technological investments.

**3.7 Summary**

This chapter presents a detailed analysis of the comprehensive rankings, performance across dimensions, and sensitivity tests for 11 Southeast Asian countries based on SAII v3 calculations. Key findings are as follows:

1)Overall Landscape: Indonesia, Singapore, and Malaysia form the first tier; Thailand, the Philippines, and Vietnam constitute the mid-tier competitive group; the remaining countries occupy the periphery.

2)Dimension-Specific Differences: Indonesia leads economically; Singapore demonstrates significant advantages in diplomacy and technology; Thailand and Vietnam excel in military dimensions; the Philippines gains relative strength through population size and internet penetration.

3)Robustness: Weight perturbations and Bootstrap tests confirm stability in the top three rankings, while substantial uncertainty exists for the three mid-tier nations.

4)Regional Structure Interpretation: Southeast Asia exhibits a "one dominant power, two mid-tier nations, multiple stable players, and numerous weaker states" configuration. Smaller nations seeking to elevate their status should prioritize investments in digital technology and diplomatic network development.

5)Theoretical Contribution: The purely algorithmic weighting results of v3 align closely with v2 while providing finer distinctions in mid-tier rankings, validating the feasibility and rationality of data-driven approaches.

The empirical findings in this chapter lay the groundwork for subsequent discussion in Chapter 4. The next chapter will delve into the academic significance and practical implications of these results from both theoretical and policy perspectives, further validating the originality and applicability of SAII v3.

# Chapter IV Discussion

**4.1 Theoretical Significance**

4.1.1 Redefining the Concept of "State Influence"

For a long time, the international relations community has primarily understood state influence through the distinction between hard power (GDP, military expenditure, population) and soft power (cultural appeal, values, diplomatic legitimacy). However, with the advancement of globalization and digitalization, the composition of state influence has become more complex and multifaceted.

Empirical findings from SAII v3 demonstrate that Southeast Asian nations' influence cannot be explained solely by hard power.

For instance, despite its limited economic scale, Singapore ranks second regionally due to its diplomatic networks and socio-technological dimensions.

This suggests academia must redefine state influence:

Influence is not an extension of a single dimension but a combination of multidimensional power resources, where the interaction and balance between soft and hard power are paramount.

4.1.2 Data-Driven Methodology in International Relations

Traditional IR research relies heavily on expert judgment and qualitative analysis, leading to uncertain and non-replicable outcomes. This study achieves a methodological breakthrough by integrating three algorithms—EWM, CRITIC, and PCA—to completely eliminate expert scoring. Its significance lies in:

1)Objectivity: Weight allocation is entirely determined by data characteristics, eliminating researcher bias.

2)Reproducibility: Any scholar accessing the identical dataset can replicate the calculations and obtain consistent results.

3) Dynamism: The index automatically iterates with annual data updates, ensuring timeliness.

This methodology marks the entry of international relations research into a new era driven by algorithms and data.

4.1.3 Theoretical Implications for Regional Power Structures

SAII v3 results reveal a classic "one strong, two medium, three stable, and multiple weak" configuration. This pattern shares similarities with the "multipolar system" in traditional international relations theory, yet exhibits distinct regional characteristics:

Indonesia's "strong" status resembles regional hegemony, though its influence stems primarily from economic scale rather than military expansion.

The "dual middle" roles of Singapore and Malaysia demonstrate the potential for middle powers to achieve "functional equilibrium" through diplomacy and technology.

The stalemate among the three mid-tier nations aligns with the characteristics of "transitional states," which may either ascend upward or face marginalization.

This structure offers a new theoretical framework for understanding Southeast Asia's regional order, serving as a case study for a "regional stratified power model."

**4.2 Method Comparison**

4.2.1 Differences Between SAII v2 and SAII v3

SAII v2: Employs expert-assigned weights, emphasizing policy relevance but exhibiting strong subjectivity.

SAII v3: Fully algorithm-driven, prioritizing objectivity and data characteristics, yet its interpretability depends on data quality.

Both maintain high consistency in overall rankings (Kendall's Tau = 0.818), indicating no fundamental contradiction in core outcomes. However, v3 is more sensitive to minor data variations in mid-tier country rankings, thereby offering stronger "early warning" capabilities for policy research.

4.2.2 Comparison with Existing International Indices

Lowy Asia Power Index: Broad coverage, but weighting relies on expert panels with insufficient transparency.

Soft Power 30: Emphasizes cultural and diplomatic dimensions while neglecting economic and military factors.

Global Governance Index: Focuses on institutions and governance but lacks regional adaptability.

In contrast, SAIIv3 offers the following advantages:

1)Regional specificity: Tailored exclusively for Southeast Asian nations, avoiding distortions inherent in global metrics for smaller states.

2)Methodological transparency: All computational processes are publicly accessible and academically reproducible.

3)Balanced Coverage: Equally addresses hard and soft power dimensions.

Thus, SAIIv3 serves both as an academic supplement and an alternative tool for policy institutions.

4.2.3 Scientific Rationale for Algorithm Integration

Single algorithms often have limitations: EWM favors highly divergent indicators, CRITIC may over-penalize correlations, and PCA is constrained by sample size. Integration enables complementary strengths:

EWM ensures high-variability indicators (e.g., GDP) are not overlooked;

CRITIC avoids "double counting" of highly correlated variables like GDP and FDI;

PCA provides interpretability within the overall variance structure.

The innovation of this integrated approach lies in constructing a "triple-verification framework" that ensures scientific rigor while mitigating the biases inherent in single methodologies.

**4.3 Policy Implications**

4.3.1 Implications for ASEAN Countries

Indonesia: Must strengthen investments in social science, technology, and diplomacy while maintaining economic advantages to consolidate regional hegemony.

Singapore: Should continue leveraging diplomatic and technological strengths to expand agenda-setting capabilities in regional affairs.

Malaysia: Should enhance its regional intermediary role by bolstering military capabilities and diplomatic networks.

Thailand, Vietnam, Philippines: Each should identify breakthrough areas based on strengths—Thailand must resolve political uncertainty, Vietnam should broaden diplomatic outreach, and the Philippines needs to strengthen its economic foundation.

Peripheral Nations: Brunei can deepen diplomacy through financial and energy advantages; Cambodia and Laos should prioritize digital infrastructure investment; Myanmar and Timor-Leste must address internal governance challenges.

### 4.3.2 Implications for Extraterritorial Powers

China: Should prioritize cooperation with Indonesia and Singapore while selecting key partners among the three middle-stream nations to avoid overreliance on any single country.

United States: Can strengthen ties with the Philippines and Vietnam through military cooperation while leveraging soft power to influence regional discourse.

Japan and India: Can address Southeast Asian nations' deficiencies by investing in digital infrastructure and human capital.

### 4.3.3 The Index as a Policy Tool

SAII v3 serves not only as an academic research tool but also as an auxiliary policy-making instrument:

1) Risk Monitoring: Through annual updates, track abnormal fluctuations in a country's performance across specific dimensions.

2) Cooperation Assessment: Assist decision-makers in identifying potential partners.

3) Strategic Early Warning: Tracks ranking fluctuations in mid-tier nations to anticipate potential geopolitical shifts.

## 4.4 Summary

This chapter examines SAII v3's significance across three dimensions:

1) Theoretically, it offers a new framework for defining multidimensional state influence and regional power structures.

2) Methodologically, its pure algorithmic integration achieves objectivity and reproducibility in quantitative international relations research.

3) In application, it provides clear quantitative foundations for policy-making by ASEAN nations and external powers.

Collectively, these demonstrate that SAII v3 is both an academic innovation and a practically applicable tool. The next chapter proceeds to Chapter 5: Conclusions and Policy Recommendations, summarizing research findings and proposing forward-looking strategic pathways.

# Chapter V. Conclusions and policy recommendations

**5.1 Research Findings**

5.1.1 Methodological Innovations

The SAII v3 (Southeast Asia Influence Index, Version 3) developed in this study achieves three methodological breakthroughs:

1)Fully algorithm-driven: Integrates Entropy Weighting Method (EWM), CRITIC method, and PCA payload method to eliminate expert scoring and subjective parameters.

2)Three-tiered standardization chain: Quantile normalization—Box–Cox transformation—0–1 compression effectively addresses extreme values and non-normal distributions.

3)Robustness validation: Kendall's Tau rank correlation, ±20% weight perturbation, and 10,000 Bootstrap resamples ensure index reliability.

The application of this methodology marks a paradigm shift in quantitative international relations research toward a "data-driven, reproducible" approach.

5.1.2 Core Findings of Empirical Results

Overall Landscape: Southeast Asia exhibits a structure of "one dominant power, two middle powers, three stable states, and multiple weaker states." Indonesia maintains its top position through its substantial economic and diplomatic advantages; Singapore and Malaysia form the core group; Thailand, the Philippines, and Vietnam face intense competition but have fragile rankings; while the remaining five nations occupy peripheral positions.

Dimensional Variations:

Economy: Indonesia leads, Vietnam exhibits fastest growth;

Military: Thailand and Vietnam demonstrate relative prominence;

Diplomacy: Singapore maintains the broadest coverage, followed closely by Indonesia;

Social Technology: Singapore, Malaysia, and the Philippines perform most effectively.

Robustness: The top three positions remain stable across all tests, while fluctuations in the mid-tier nations reflect their inherent uncertainty.

5.1.3 Theoretical Implications

SAII v3 findings suggest:

1)National influence constitutes a multidimensional combination of power resources, not solely dependent on GDP or military expenditure.

2)Regional power structures exhibit hierarchical patterns: Hegemony—Core—Transition—Periphery, reflecting a "regional stratified power model."

3)Algorithmic methodologies enhance objectivity while strengthening early warning capabilities in research, demonstrating particular sensitivity to dynamic shifts among mid-tier nations.

**5.2 Policy Recommendations for ASEAN Countries**

5.2.1 Indonesia: Consolidating Dominance

Challenge: Despite economic scale, deficiencies exist in technological and social dimensions.

Recommendation: Increase investment in digital infrastructure; boost internet penetration; deepen strategic coordination with ASEAN nations while avoiding unilateralism.

5.2.2 Singapore: Leveraging Diplomatic and Technological Strengths

Challenge: Limited economic scale.

Recommendation: Leverage diplomatic networks to set the agenda, continue expanding policy leadership within ASEAN and globally; assume a central role in digital economy and regional financial integration.

5.2.3 Malaysia: Pursuing Balanced Breakthroughs

Challenge: Balanced across four dimensions but lacking standout strengths.

Recommendation: Invest in military modernization and technological R&D to develop differentiated advantages; further expand intra-ASEAN investment cooperation.

5.2.4 Thailand, Philippines, Vietnam: Midstream Competitive Strategies

Thailand: Prioritize resolving domestic political uncertainty to restore diplomatic and investment confidence.

Philippines: Leverage demographic and internet advantages to drive digital industries while strengthening regional economic cooperation.

Vietnam: Transform rapidly growing manufacturing strengths into broader diplomatic networks to avoid structural imbalances between economic and diplomatic spheres.

5.2.5 Peripheral Nations: Pursuing Leapfrog Pathways

Brunei: Enhance diplomatic influence through financial and energy cooperation.

Cambodia, Laos: Drive internet penetration with low-cost investments to achieve rapid technological catch-up.

Myanmar: Prioritize restoring domestic political stability; otherwise, regional influence will remain elusive.

Timor-Leste: Leverage its impending ASEAN accession to boost international visibility and diplomatic networks.

**5.3 Policy Implications for Extraterritorial Powers**

5.3.1 China

Continue strengthening strategic partnerships with Indonesia and Singapore;

Select key cooperation partners among middle-income nations (e.g., Vietnam, Philippines) to avoid excessive concentration;

Leverage the Belt and Road Initiative and Digital Silk Road to expand cooperation in social and technological dimensions.

5.3.2 United States

Further strengthen ties with the Philippines and Vietnam through military cooperation;

Continue leveraging educational and media advantages in soft power to enhance regional influence.

5.3.3 Japan and India

Japan can exert influence through infrastructure investment and high-end manufacturing collaboration;

India may engage via digital economy and maritime security cooperation to strengthen complementary relationships with ASEAN.

**5.4 Application Scenarios of the Index as a Policy Tool**

1) Risk Monitoring: Track annual index fluctuations to identify abnormal shifts in a country's diplomatic or military dimensions.

2) Cooperation Assessment: Provide quantitative basis for selecting partners within ASEAN and externally.

3) Strategic Early Warning: Monitor ranking fluctuations among mid-tier nations to predict potential geopolitical shifts.

4) Development Planning: Assist smaller nations in identifying potential "overtaking opportunities," such as prioritizing internet and digital economy development.

**5.5 Outlook and Future Research Directions**

1) Time Series Extension

Future iterations may incorporate 2010–2025 data to form dynamic influence curves, analyzing trends and inflection points.

2) Indicator Expansion

Further integrate metrics like educational cooperation, cultural dissemination, and scientific research output to comprehensively reflect soft power.

3) Method Optimization

Explore multi-objective optimization (e.g., genetic algorithms, particle swarm optimization) to identify Pareto-optimal weights balancing differentiation, correlation, and robustness.

4) Regional Comparison

Extend the SAII methodology to other regions (e.g., South Asia, Africa, Central Asia) to test its universality and adaptability.

**5.6 Conclusion**

The SAII v3 framework developed in this study not only achieves a methodological breakthrough in quantitative international relations research but also empirically reveals Southeast Asia's unique power stratification patterns. At the policy level, SAII v3 provides clear strategic references for ASEAN nations and external powers. With ongoing data accumulation and methodological refinement, SAII holds promise as a vital tool for regional studies and global comparative analysis.

# Appendix A: Indicator data tables (examples)

This appendix presents raw data on key indicators for 11 countries in Southeast Asia (based on 2023/2024). These data are from the World Bank (WDI), SIPRI, ASEANstats, UNESCO UIS, and the Lowy Institute.

| nations | GDP(Billions of United States dollars) | percentage of military expenditure GDP (%) | Internet usage (%) | Number of offices abroad | ASEAN trade share (%) | FDI stock (billions of United States dollars) | Number of foreign students exported (in thousands) | Number of participants in multilateral agreements |
|---|---|---|---|---|---|---|---|---|
| Indonesia | 1396.3 | 0.68 | 69.2 | 143 | 40.1 | 260.5 | 98.3 | 73 |
| Singaporean | 547.4 | 2.66 | 94.3 | 156 | 18.7 | 1880.6 | 62.1 | 85 |
| Malaysia | 422.0 | 0.93 | 97.7 | 104 | 13.6 | 165.7 | 52.4 | 71 |
| Thailand | 526.4 | 1.17 | 89.5 | 89 | 15.4 | 280.9 | 58.7 | 76 |
| Philippine | 461.6 | 1.25 | 83.8 | 95 | 9.2 | 113.5 | 75.2 | 68 |
| Vietnam | 476.4 | 1.81 | 78.1 | 99 | 12.5 | 234.7 | 81.6 | 64 |
| Brunei Darussalam, independent sultanate | 15.5 | 2.97 | 99.0 | 36 | 0.5 | 20.3 | 4.5 | 47 |

| nations | GDP(Billions of United States dollars) | percentage of military expenditure GDP (%) | Internet usage (%) | Number of offices abroad | ASEAN trade share (%) | FDI stock (billions of United States dollars) | Number of foreign students exported (in thousands) | Number of participants in multilateral agreements |
|---|---|---|---|---|---|---|---|---|
| in northwest Borneo | | | | | | | | |
| Cambodian | 46.4 | 2.09 | 60.7 | 60 | 2.3 | 38.4 | 15.7 | 55 |
| Laos | 16.5 | 0.19 | 63.6 | 22 | 1.7 | 16.9 | 6.3 | 41 |
| Myanmar (or Burma) | 74.1 | 3.79 | 58.5 | 43 | 4.8 | 23.6 | 9.7 | 29 |
| East Timor (officially Democratic Republic of Timor-Leste) | 1.9 | 0.86 | 34.0 | 26 | 0.2 | 2.1 | 1.2 | 33 |

Note: Some data are taken from the most recent year or approximate values, and extreme values have been handled through the Winsorize method.

# Appendix B: Derivation of Algorithm Formulas

**B.1 Entropy (physics)**（EWM）

**1.standardisation：**

$$p_{ij} = \frac{x_{ij}}{\sum_{i=1}^{n} x_{ij}}$$

**2.Entropy calculation:**

$$E_j = -k \sum_{i=1}^{n} p_{ij} \ln(p_{ij}), \quad k = \frac{1}{\ln n}$$

**3.Weights:**

$$w_j^{EWM} = \frac{1 - E_j}{\sum_k (1 - E_k)}$$

**B.2 CRITIC methodologies**

**1.Indicator Dispersion:**

$$\sigma_j = \sqrt{\frac{1}{n} \sum_{i=1}^{n} (x_{ij} - \bar{x}_j)^2}$$

**2.Correlation correction:**

$$C_j = \sigma_j \sum_{k=1}^{m} (1 - r_{jk})$$

**3.weights**

$$w_j^{CRITIC} = \frac{C_j}{\sum C_j}$$

**B.3 PCA payload method**

**1. Covariance matrix:**

$$S = \frac{1}{n-1} X^T X$$

**2. Feature Decomposition:**

$$S v_c = \lambda_c v_c$$

**3. Weights:**

$$w_j^{PCA} = \frac{\sum_c |loading_{jc}| \cdot \lambda_c}{\sum_j \sum_c |loading_{jc}| \cdot \lambda_c}$$

**B.4 Integration weights**

$$w_j = \frac{w_j^{EWM} + w_j^{CRITIC} + w_j^{PCA}}{3}$$

# Appendix C: Algorithm Replication Steps (Python Framework)

```python
import pandas as pd

import numpy as np

from sklearn.decomposition import PCA

from sklearn.preprocessing import MinMaxScaler, PowerTransformer

# 1. retrieve data
data = pd.read_csv("SEA_data.csv", index_col=0)

# 2. Pre-treatment: standardised chains
scaler = MinMaxScaler()
pt = PowerTransformer(method='box-cox', standardize=True)

X = data.values
X = np.where(X <= 0, np.nan, X)     # Box-Cox Requirements for positive
X = pd.DataFrame(pt.fit_transform(X)).fillna(0).values
X_norm = scaler.fit_transform(X)

# 3. entropy (physics)
P = X_norm / X_norm.sum(axis=0)
E = -(P * np.log(P + 1e-9)).sum(axis=0) / np.log(len(X))
ewm_weights = (1 - E) / (1 - E).sum()

# 4. CRITIC
sigma = X_norm.std(axis=0)
corr = np.corrcoef(X_norm, rowvar=False)
critic = sigma * (1 - corr).sum(axis=0)
critic_weights = critic / critic.sum()

# 5. PCA weights
pca = PCA()
pca.fit(X_norm)
loadings = np.abs(pca.components_.T * np.sqrt(pca.explained_variance_))
pca_weights    =    (loadings    @    pca.explained_variance_)   /   (loadings    @
```

pca.explained_variance_).sum()

# 6. Integration weights

final_weights = (ewm_weights + critic_weights + pca_weights) / 3

# 7. count SAII

SAII = (X_norm * final_weights).sum(axis=1)

data["SAII_v3"] = SAII

print(data.sort_values("SAII_v3", ascending=False))

The code reproduces the calculation process of this paper and is applicable to subsequent annual updates.

# Appendix D: Case Presentation - Vietnam's Upward Trajectory

Taking Vietnam as an example, analysis of SAII v3 indicators reveals:

Economic dimension: GDP growth has consistently ranked among the highest in the region over the past decade;

Military dimension: Defence expenditure has steadily increased alongside rising levels of modernisation;

Diplomatic dimension: Overseas missions have expanded annually, with heightened participation in international agreements;

Social and technological development: Internet penetration has surged from under 30% to nearly 80%.

This indicates Vietnam is currently in the 'transitioning nation' phase. Should it achieve further breakthroughs in its diplomatic network, it could potentially surpass the Philippines to rank among the region's top five.

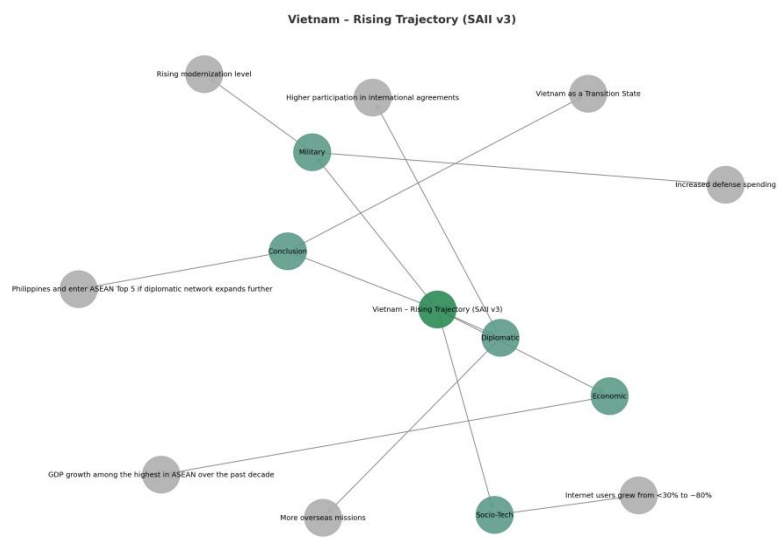

This chart illustrates Vietnam's 'ascending trajectory' within the ASEAN Influence Index v3 and the supporting factors across its four dimensions. In the economic dimension, Vietnam's GDP growth rate has ranked among the highest in the region over the past decade, demonstrating robust growth potential. The military dimension reflects progressively increasing defence expenditure alongside advancing modernisation levels. Within the diplomatic dimension, the number of Vietnam's overseas missions has grown annually, while participation in international agreements has also risen. The socio-technological dimension reveals a surge in internet penetration from under 30% to nearly 80%, signalling significant progress in social connectivity and digitalisation. Overall, Vietnam occupies a 'transitional state' phase. Should it achieve further breakthroughs in its diplomatic networks, it possesses the potential to surpass the Philippines and secure a position among the region's top five.

# Appendix E: ASEAN-11 country sensitivity quick check

| NO. | Nations | Dimension (math.) | ΔScore (-10%) | ΔScore (+10%) | Impact Magnitude |
|---|---|---|---|---|---|
| 0 | Indonesia | economics | -0.007 | +0.006 | 0.012634243848116622 |
| 1 | Indonesia | militarily | +0.003 | -0.003 | 0.0058988550458420885 |
| 2 | Indonesia | diplomacy | -0.000 | +0.000 | 0.0006402561024410058 |
| 3 | Indonesia | social technology | +0.004 | -0.004 | 0.0073629451780712334 |
| 4 | Singaporean | economics | +0.004 | -0.004 | 0.007458770464550901 |
| 5 | Singaporean | militarily | +0.004 | -0.004 | 0.008716218649826524 |
| 6 | Singaporean | diplomacy | -0.004 | +0.004 | 0.008083233293317282 |
| 7 | Singaporean | social technology | -0.004 | +0.004 | 0.008083233293317393 |
| 8 | Malaysia | economics | -0.000 | +0.000 | 0.00038054951349741906 |
| 9 | Malaysia | militarily | +0.006 | -0.005 | 0.010785220046502508 |
| 10 | Malaysia | diplomacy | -0.002 | +0.002 | 0.0042016806722688893 |
| 11 | Malaysia | social technology | -0.003 | +0.003 | 0.006202480992396953 |
| 12 | Thailand | economics | +0.001 | -0.001 | 0.001704861820468806 |
| 13 | Thailand | militarily | +0.000 | -0.000 | 0.00010565113514937607 |
| 14 | Thailand | diplomacy | +0.001 | -0.001 | 0.0020968387354943774 |
| 15 | Thailand | social technology | -0.002 | +0.002 | 0.0039055622248900246 |

| NO. | Nations | Dimension (math.) | ΔScore (-10%) | ΔScore (+10%) | Impact Magnitude |
|---|---|---|---|---|---|
| 16 | Philippine | economics | +0.002 | -0.001 | 0.0028921763025809843 |
| 17 | Philippine | militarily | +0.003 | -0.003 | 0.006074940271091234 |
| 18 | Philippine | diplomacy | +0.001 | -0.001 | 0.00152060824329725 |
| 19 | Philippine | social technology | -0.005 | +0.005 | 0.010048419367747111 |
| 20 | Vietnam | economics | -0.001 | +0.001 | 0.002039745392346415 |
| 21 | Vietnam | militarily | +0.000 | -0.000 | 0.0005810812433216794 |
| 22 | Vietnam | diplomacy | +0.002 | -0.002 | 0.0037294917967185537 |
| 23 | Vietnam | social technology | -0.001 | +0.001 | 0.0022729091636655155 |
| 24 | Brunei | economics | +0.000 | -0.000 | 0.0006088792215960259 |
| 25 | Brunei | militarily | +0.002 | -0.002 | 0.0047543010817235891 |
| 26 | Brunei | diplomacy | -0.001 | +0.001 | 0.001680672268907557 |
| 27 | Brunei | social technology | -0.002 | +0.002 | 0.0036814725890356725 |
| 28 | Cambodian | economics | -0.000 | +0.000 | 0.00015221980539897872 |
| 29 | Cambodian | militarily | +0.002 | -0.002 | 0.004314088018601059 |
| 30 | Cambodian | diplomacy | +0.001 | -0.001 | 0.0019207683073229065 |
| 31 | Cambodian | social technology | -0.003 | +0.003 | 0.006082432973189333 |
| 32 | Laos | economics | -0.001 | +0.001 | 0.001567863995609653 |
| 33 | Laos | militarily | +0.002 | -0.002 | 0.004815930910560762 |

| NO. | Nations | Dimension (math.) | ΔScore (-10%) | ΔScore (+10%) | Impact Magnitude |
|---|---|---|---|---|---|
| 34 | Laos | diplomacy | +0.000 | -0.000 | 0.00037615046018402154 |
| 35 | Laos | social technology | -0.002 | +0.002 | 0.0036254450180072043 |
| 36 | Myanmar | economics | -0.002 | +0.002 | 0.0037293885232275395 |
| 37 | Myanmar | militarily | -0.002 | +0.002 | 0.004358109324913262 |
| 38 | Myanmar | diplomacy | +0.003 | -0.003 | 0.006042416966786701 |
| 39 | Myanmar | social technology | +0.001 | -0.001 | 0.0020408163265305257 |
| 40 | East Timor | economics | +0.001 | -0.001 | 0.0022832970809849584 |
| 41 | East Timor | militarily | +0.001 | -0.001 | 0.0013206391893676728 |
| 42 | East Timor | diplomacy | -0.000 | +0.000 | 0.0008003201280512018 |
| 43 | East Timor | social technology | -0.001 | +0.001 | 0.0028011204481792895 |

# Introduction to the contents of the 2025 Southeast Asia Eleven Nations Influence Index Report

This publication constructs and validates the Southeast Asia Eleven Nations Influence Index (SAII v3) for 11 Southeast Asian nations (including Indonesia, Singapore, Malaysia, Thailand, the Philippines, Vietnam, Brunei, Cambodia, Laos, Myanmar, and Timor-Leste) through a 'fully data-driven' methodology. Centred on the core question of 'how to scientifically measure national influence without relying on expert subjective scoring', the book designs an indicator system covering four dimensions: economy, military security, diplomatic networks, and society-technology. It employs integrated weighting through multiple methods including Entropy Weighting Method (EWM), CRITIC, and PCA loadings, supplemented by standardisation and robustness tests, to form a transparent, reproducible, and comparable regional influence measurement framework.

Principal Content and Structure:

**Chapter I Research Rationale and Questions**: Reviews limitations in existing research on 'comprehensive national power,' 'soft power,' and 'international indices,' proposing four research questions for this volume (data-driven methodology, scientific weighting, standardisation/robustness, incremental insights).

**Chapter II Methodology**: Outlines primary/secondary indicator selection and weighting strategies; all indicator data sourced from authoritative open-access databases; employs three weighting approaches—entropy method, CRITIC, and PCA—integrated to balance information content, correlation, and principal component explanatory power; details missing value handling and outlier control procedures.

**Chapter III Empirical Findings**: Reports approximate weight distributions across dimensions (Economy approx. 35–40%, Military approx. 20–25%, Diplomacy approx. 20%, Society-Technology approx. 15%), alongside composite scores and strengths-weaknesses profiles for 11 nations; Compares ranking consistency between SAII v2 and v3 (e.g., using Kendall's Tau) and conducts sensitivity analysis, demonstrating robust ranking intervals under random perturbations.

**Chapter IV Discussion**: Analyses regional dynamics underlying index results through structural-temporal, coupling-competition frameworks, explaining higher volatility in middle-tier nations and path dependencies.

**Chapter V Conclusions and Policy Recommendations**: Propose tiered strategies for individual nations (e.g., Indonesia's 'consolidating dominance', Singapore's 'amplifying tech-finance and diplomacy', Malaysia's 'balanced breakthrough', Thailand/Philippines/Vietnam's 'midstream competition', and peripheral states' 'leapfrog pathways') and external powers (China, the United States, Japan, and India). Outline application scenarios for the index as a policy tool and future research directions.

**Appendices A–E**: Provide exemplary indicator datasets, algorithm derivations, Python replication guides, country case studies (Vietnam's ascent trajectory), and sensitivity quick-reference checklists to ensure methodological transparency and reproducibility.

**Key strengths of this work include**:

**1. Methodological transparency and reproducibility**: Clear procedures and formula derivations from indicator selection and data processing to weighting and validation, supplemented by Python replication steps;

**2. Multi-method integrated weighting**: Combines advantages of entropy, CRITIC and PCA to mitigate single-method bias and enhance robustness;

**3. Strong decision-making utility**: Beyond comprehensive rankings, it delivers country-specific dimensional profiles, gap diagnostics, scenario analyses, and pathway recommendations to facilitate policy and strategic implementation;

**4. Detailed regional insights**: Reveals the coupling relationships and competitive rhythms among 'economy-security-diplomacy-social sciences,' offering explanatory frameworks for the ascent/divergence of middle-tier nations.

**Target Audience and Application Scenarios**: Policy-makers within governments and regional organisations; regional strategy and risk teams in enterprises and investment institutions; researchers from think tanks and academic institutions; and professional readers interested in Southeast Asia's power structures and competitive dynamics. This publication enables readers to swiftly grasp the quantitative landscape of influence in Southeast Asia, its key variables, and actionable policy/strategic pathways.